# WINDOWS AND LINUX OPERATING SYSTEMS FROM A SECURITY PERSPECTIVE

Youssef Bassil

LACSC – Lebanese Association for Computational Sciences
Registered under No. 957, 2011, Beirut, Lebanon
youssef.bassil@lacsc.org

*Abstract*: Operating systems are vital system software that, without them, humans would not be able to manage and use computer systems. In essence, an operating system is a collection of software programs whose role is to manage computer resources and provide an interface for client applications to interact with the different computer hardware. Most of the commercial operating systems available today on the market have buggy code and they exhibit security flaws and vulnerabilities. In effect, building a trusted operating system that can mostly resist attacks and provide a secure computing environment to protect the important assets of a computer is the goal of every operating system manufacturer. This paper deeply investigates the various security features of the two most widespread and successful operating systems, Microsoft Windows and Linux. The different security features, designs, and components of the two systems are to be covered elaborately, pin-pointing the key similarities and differences between them. In due course, a head-to-head comparison is to be drawn for each security aspect, exposing the advantage of one system over the other.

## INTRODUCTION

An operating system, also called OS, is a collection of system programs, tools, and utilities that manage computer hardware resources and offer common services for client application software [1]. The operating system is the first program to execute upon booting a computer and is thus considered the most vital type of system software. An operating system runs users' application programs and provides them a suitable interface to interact with the computer hardware. It is additionally responsible for carrying out other tasks including but not limited to spawning processes, creating threads, allocating primary memory to various applications, managing data storage, controlling input and output peripherals, hosting device drivers, and delivering multi-level secure execution platform. Microsoft Windows and Linux are two of the most renowned operating systems that have a widespread use in every computer-related field.

Microsoft Windows is a proprietary operating system that targets the Intel-based PC architectures. Windows including all its versions is estimated to have 92.03% total net market share [2], making it the largest dominant operating system for personal computers. Windows is designed by Microsoft Corporation who originated it in 1985 as an add-on for MS-DOS, which was the standard operating system shipped on most Intel-based PCs at the time. Today, Microsoft Windows has gone through several versions, the most recent version for personal computers is Windows 7; while, the most recent version for server computers is Windows Server 2008 R2 [3].

Conversely, Linux is a Unix-like operating system composed of a Linux kernel originally developed by Linus Torvalds and later extended and improved by a large community of developers over the world, and the GNU which is a software collection made out of software parts, system programs, and utility tools originally conceived by Richard Stallman to create a completely free and open operating system using the Linux kernel. For this reason, the joint product of the Linux kernel and the GNU software collection is more commonly called GNU/Linux. Basically, GNU/Linux is open-source and therefore anyone can read and modify its source code and create what so called Linux distributions such as Red Hat, Debian, Ubuntu, SuSE, and Google Android [4].

In a computer security context, almost any operating system including Windows and Linux are faced with security vulnerabilities, bugs, and flaws throughout their lifetime [5]; nevertheless, operating systems makers endeavor to regularly solve all type of security imperfections in their products so as to deliver the most possible secure computing environment for computer users and their application programs. In effect, a trusted operating system is an operating system that provides a reliable security framework and a multilevel secure computing environment for both users and programs [6].

This paper investigates two operating systems from a security perspective, the Windows and the Linux OS, by describing their internal security models and shedding the light on the key differences and similarities between their security design, architecture, processes, and algorithms.

## THE SECURITY MODEL

### Windows

The Windows security model is a collection of user-mode and kernel-mode processes that deliver, monitor, and manage the different OS security components, and coordinates among them. Figure 1 depicts the Windows security model together with its components [7].

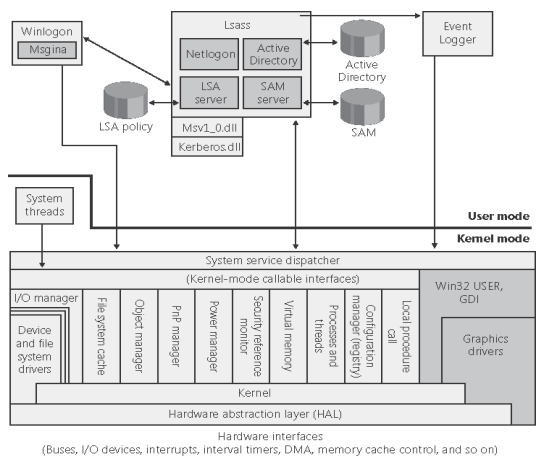

Figure 1. Windows security model

### 1. Security Reference Monitor (SRM)

SRM is a component running in kernel mode (c:\windows\system32\Ntoskrnl.exe) that enforces security policies on the local computer. It guards the various operating system resources by performing run-time object protection and auditing, and manipulating security privileges often know as user rights [7].

### 2. Local Security Authority Subsystem (Lsass)

Lsass is a user-mode process (c:\Windows\System32 \Lsass.exe) that is responsible for the local system security policy, user authentication, and sending security audit messages to the event log. Actually, Lsass implements most of its functionalities in a dynamic-link library (c:\Windows \System32\Lsasrv.dll).

### 3. Lsass Policy Database

It is a database that contains the local system security policy settings. This database is stored in the registry under HKLM\SECURITY. It includes such information as what domains are entrusted to authenticate logon attempts, who has permission to access the system and how (interactive, network, and service logons), who is assigned which privileges, and what kind of security auditing can be performed.

### 4. Security Accounts Manager (SAM)

SAM is a combination of a service and a database. The SAM service is a set of subroutines responsible for managing the database that contains the usernames and groups defined on the local machine. It is implemented as a dynamic-link library (\Windows\System32\Samsrv.dll), and runs in the Lsass process. On the other hand, the SAM database is used on systems not functioning as domain controllers and contains the defined local users and groups, along with their passwords and other attributes. The SAM database is stored in the registry under HKLM\SAM [8].

### 5. Active Directory

It is a directory service that contains a database to store information about objects in a domain. A domain is a collection of computers and their associated security groups that are managed as a single entity. The Active Directory stores information about the objects in the domain, including users, groups, computers, passwords, and privileges. The Active Directory server is implemented as \Windows\System32\Ntdsa.dll, and runs in the Lsass process [9].

### 6. Network Logon Service (Netlogon)

It is a Windows service (\Windows\System32\Netlogon.dll) that supports authentication of account logon events in a domain. It additionally verifies logon requests, and registers, authenticates, and discovers domain controllers.

### 7. Authentication Packages

They are dynamic-link libraries (DLLs) that run in the context of the Lsass process and implement the Windows authentication policy. An authentication DLL is responsible for checking whether a given username and password match, and if so, returning to Lsass the information detailing the user's security identity. The Windows authentication packages include Kerberos and MSV1_0 [7].

### 8. Logon Process (Winlogon)

It is a user-mode process (\Windows\System32\Winlogon. exe) that is responsible for responding to the Lsass and for managing interactive logon sessions. Winlogon creates a GUI user's shell process when a user logs on.

### 9. Graphical Identification and Authentication (GINA)

It is a user-mode DLL that runs in the Winlogon process and that Winlogon uses to obtain a username and password or smart card PIN. The GINA standard library is located at \Winnt\System32\Msgina.dll [10].

*Linux*

The Linux security model is a collection of several active processes, daemon services, and libraries that provide a secure framework for the Linux kernel to work in. Figure 2 depicts the Linux security model along with it various modules [11].

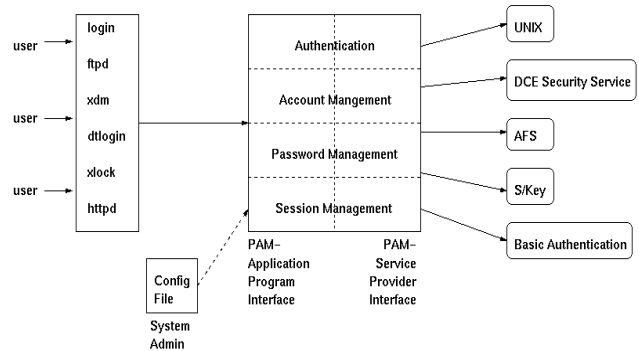

Figure 2. Linux security model

### 1. PAM Library

The Pluggable Authentication Modules (PAM) library provides the necessary interface and functions for developing PAM-aware applications. The PAM library is essential to allow the authentication of users in the Linux operating system.

### 2. PAM Configuration File

It is a text file where the system administrator can specify which authentication scheme is used for a particular application. On the Linux system, this configuration information can be stored either in a file under the /etc/pam directory or as a line in the /etc/conf configuration file. Upon the initialization of the PAM library, The PAM configuration file is read so as to load the corresponding authentication modules [12].

### 3. Authentication Module

It is a module containing several authentication procedures, used for creating authentication credentials, authenticating users, and granting privileges to authenticated users.

### 4. Account Management Module

It manages user accounts and establishes whether an authenticated user is permitted to gain access to the system. It creates login session after a successful authentication and is responsible for validating the expiration date of the username and/or password.

### 5. Password Management Module

It handles and manages users' passwords including setting, resetting, and changing passwords. In other words, it sets or changes the user's authentication data.

### 6. Session Management Module

It manages the beginning and the end of a login session. It also deals with creating the appropriate log entries for every initialized session.

*Head-to-Head Comparison*

Although both systems have their own standards and design, they are both modularized in a way that their security components are sort of independent services and processes working in the kernel mode and in the user mode. These processes are used by the operating system to accomplish a specific task such as authentication, logging, enforcing policies, and account management. Such modularization makes the system more stable and easier to be updated and extended.

## IDENTIFICATION

An ID is a method of uniquely identifying entities that perform actions in a system. Entities can be users, resources, processes, domains, LAN, etc.

*Windows*

An SID is a variable-length numeric value that consists of an SID structure revision number, a 48-bit authority ID, and a variable number of 32-bit sub-authority that compose the actual unique ID of the entity and a relative identifier (RID) value [13]. Figure 3 is an SID sample.

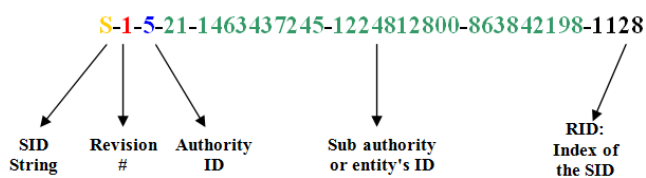

Figure 3. SID format

An SID is composed of the following elements:

- SID String: It implies that this string is an SID.
- Revision Number: It is the structure revision number.
- Authority ID: It is a number that specifies who created or granted this SID.
- Actual ID: It is the unique ID of the actual entity.
- RID: It is the relative ID, an index or ID for the SID. 1128 RID means that the system has 1128 SIDs already created.

Every user, group, and network device, as well as the login session has a unique SID. The Winlogon process is responsible for creating a unique SID for each interactive login session. The SID for a login session is usually S-1-5-5-0, with a randomly generated number for RID.

*Linux*

A user is identified by a username, which is given when the user logs on to the system. Internally, a user is identified with a User Identification Number (UID), which is a numeric value selected by the system administrator at the time the account is created. In most cases, selecting unique UIDs for each user is a good idea, though not strictly required. The mapping of username to UID is kept in the file /etc/passwd, and is centrally managed by NIS. The super user, also known as root, has a UID equals to 0. Every user belongs to one or more groups. A group is identified with a group identification number or GID for short [11].

*Head-to-Head Comparison*

Despite the difference in naming, both operating systems apply the concept of ID to uniquely identify an entity in terms of security context. Both systems generate IDs for the login session, users, and groups. The major difference resides in where each system stores its IDs. In Windows, SIDs are stored in the registry under HKLM\Security; whereas, in Linux, they are stored in the /etc/passwd file.

## ACCESS TOKENS

An Access Token is a data structure that identifies the security context of a process or thread.

*Windows*

In Windows operating system, the information in a token includes the SID, groups SID, privileges, and default DACL of the user account associated with the process or thread. When a user logs on successfully, the Winlogon process creates an initial token representing the user, and attaches the token to the initial processes it starts, by default, the Userinit.exe process. Because child processes by default inherit a copy of the access token from their creator, all processes in the user's session run under the same token. In other words, a copy of the access token is attached to every process and thread that executes on the user's behalf. Figure 4 depicts the access token data structure in Windows operating system [14].

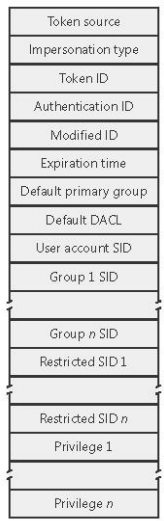

Figure 4. Windows access token

The Windows access token contains the following elements:

- The security identifier (SID) for the user's account
- SIDs for the groups of which the user is a member
- An authentication SID that identifies the current logon session
- A list of the privileges held by either the user or the user's groups
- The SID for the primary group
- The default DACL that the system uses when the user creates a securable object without specifying a security descriptor
- The source of the access token
- Whether the token is a primary or impersonation token
- An optional list of restricting SIDs
- Other statistics

*Linux*

In the Linux operating system, access tokens are data objects stored in memory and attached whenever a new process is spawned. The session management component handles the creation and attachment of access token when a new process or thread is created [15]. Figure 5 represents the different elements of an access token in Linux-based systems.

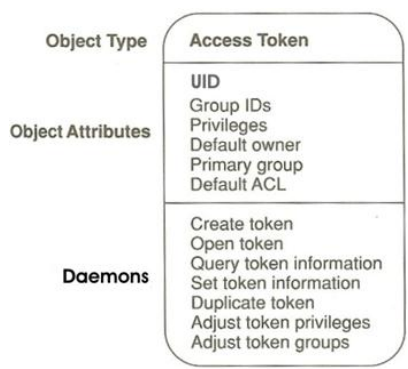

Figure 5. Linux access token

The Linux access token contains the following elements:

- The UID for the user's account
- Groups IDs represent the UIDs of the groups in which the user is a member of
- A list of the privileges and user rights
- The UID for the primary group
- The DACL contains entries that dictate who is allowed to access what

*Head-to-Head Comparison*

Both Windows and Linux use the concept of access token though each one has a different approach for implementing it. The distinct difference is that unlike Windows which stores restrictions in the access token, Linux uses DAC and MAC to impose restrictions on a particular process. Linux's access token has no restrictions entries as for the case of Windows. Moreover, Linux does not store the type of the access token (primary or impersonate) inside the token itself; rather, according to the UID, the system can deduce if this token is primary or impersonate type.

**IMPERSONATION**

Impersonation is a security concept which allows a server application to temporarily "be" the client in terms of access to secure objects [16].

*Windows*

Windows uses impersonation in its client/server programming model. For example, a server application can export resources such as files, printers, or databases. Clients wanting to access a resource send a request to the server. When the server receives the request, it must ensure that the client has permission to perform the desired operations on the resource. For example, if a user on a remote machine tries to delete a file on an NTFS share, the server exporting the share must determine whether the user is allowed to delete the file [12].

Impersonation lets a server notify the SRM that the server is temporarily adopting the security profile of a client making a resource request. The server can then access resources on behalf of the client, and the SRM can carry out the access validations. Figure 6 illustrates the impersonation mechanism in Windows. First client 1 has the right to access file x. Therefore the server upon receiving request from client 1 impersonates client 1 (substituting server's access token by the client 1's access token). Now the server through his impersonated access token can recognize that client 1 has the right to access file x and thus permission is granted and server accesses file x.

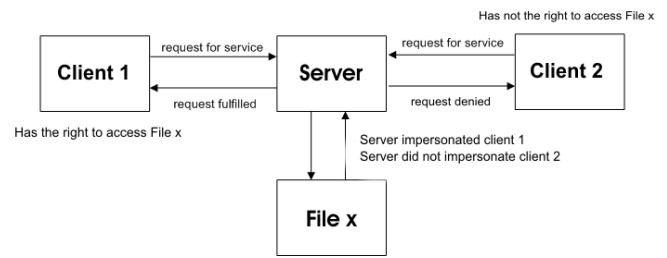

Figure 6. Windows impersonation process

*Linux*

Two separate but similar mechanisms handle impersonation in Linux, the so called set-UID, (SUID), and set-GID (SGID) mechanisms. In fact, every executable file can be marked for SUID/SGID execution. It is then executed with the permissions of the owner/group of the file, instead of the current user. Typically, certain services that require super user privileges are wrapped in a SUID-super user program, and the users of the system are given permission to execute this program. If the program can be subverted into performing some action that it was not originally intended to perform, serious breaches of security can result [13].

*Head-to-Head Comparison*

The design of impersonation in both systems is totally different. In Windows, a server can substitute its own access token by the access token of the client, then the server can decide whether the client has the right to access a particular file or not. However, in Linux, a client executes in the security context of the server whether or not that client has the right to perform a given operation. Therefore if a client in Linux has not the right to access the disk and is connected to a server that has full privileges, the client can easily access the disk through the server and that may lead to severe breaches in security.

**ACCESS CONTROL LIST**

ACL which stands for Access Control List is a list of permissions attached to an object that dictates who can access what and the level of this access, which is more commonly known as authorization [17].

*Windows*

There are two types of ACLs in Windows: DACLs and SACLs. DACL (Discretionary Access Control List) is a list of allow and deny ACE (Access Control Entries) whereas SACL (System Access Control List) specifies which operations should be logged in the security audit log [7]. Figure 7 is a representation of DACL attached to a file object.

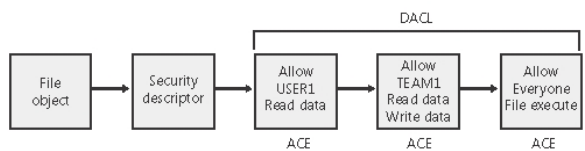

Figure 7. Windows DACL

In a DACL, each ACE contains an SID and an access mask. Four types of ACEs can appear in a DACL: access allowed, access denied, allowed-object, and denied-object. The access-allowed ACE grants access to a user and the access-denied ACE denies the access rights specified in the access mask.

In contrast, a SACL contains two types of ACEs: System audit ACEs and system audit-object ACEs. These ACEs specify which operations performed on the object by specific users or groups should be audited. Audit information is stored in the system Audit Log. Both successful and unsuccessful attempts can be audited. Figure 8 is an example of access validation. It is obvious that user DaveC is allowed to read and write to the object while group writers is denied from reading and writing to the same object [7].

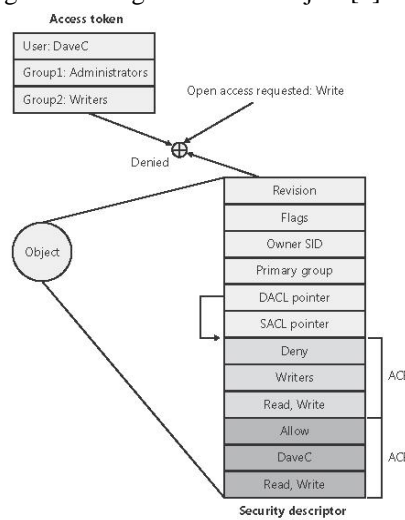

Figure 8. Windows DACL

*Linux*

Linux's access control is implemented through the file system. Each file or directory has a number of attributes, including a filename, permission bits, a UID, and a GID. The UID of a file specifies its owner. The permission bits are used to specify permissions to read (r), write (w), and execute (x) the file for the user, for the members of the user's group, and for all other users in the system. For instance, permission such as "rwxr-x-x" specifies that the owner may read, write, and execute the file; while, the group members are only allowed to read and execute it, and all others can only execute the file. A dash "-" in the permission set indicates that the access rights are disallowed. Most Linux systems today also support some form of ACL schemes.

Furthermore, each process in Linux has an effective and a real UID, as well as, an effective and a real GID associated with it. Whenever a process attempts to access a file, the kernel will use the process's effective UID and GID to compare them with the UID and the GID associated with the file to decide whether or not to grant the request.

There are two types of ACLs in Linux: DAC and MAC. DAC short for Discretionary Access Control is well at the discretion of the user. An object's owner, who is usually also the object's creator, has discretionary authority over who else may access that object. In other words, access rights are administered by the owner. In contrast, MAC short for Mandatory Access Control involves several aspects that the user cannot control or is not usually allowed to control. Objects are tagged with labels representing the sensitivity of the information contained within. MAC restricts access to objects based on their sensitivity. Subjects need formal clearance or authorization to access objects [15].

*Head-to-Head Comparison*

Both Windows and Linux implement the concept of Access Control List; nevertheless, some differences exist between the two designs. Windows uses privileges and restrictions in order to enforce system policies such as denying a user from deleting or reading a system file; whereas, Linux uses Mandatory Access Control or MAC to restrict access to system objects. Likewise, Windows uses what it calls System

Access Control List or SACL to specify which operation on a given object should be audited or logged; whereas, this concept is not implemented under Linux. Logging under Linux is done by an independent separate component.

## PRIVILEGES AND USER RIGHTS

A privilege is the authority to perform an operation that affects an entire computer rather than a particular object. User rights, also known as privileges, are assigned by administrators to individual users or groups as part of the security settings of the operating system.

### Windows

In Windows, a privilege is the right of an account to perform a particular system-related operation, such as shutting down the computer or changing the system time or accessing the registry. An account right grants or denies the account to which it is assigned. User rights are always validated in response to logon requests. For this purpose, the Local Security Authority (LSA) retrieves account rights assigned to a given user from the LSA policy database at the time the user attempts to log on to the system [14]. Figure 9 shows user rights assignment in the local security policy editor (secpol.msc), which displays the complete list of privileges and account rights available for a particular user account.

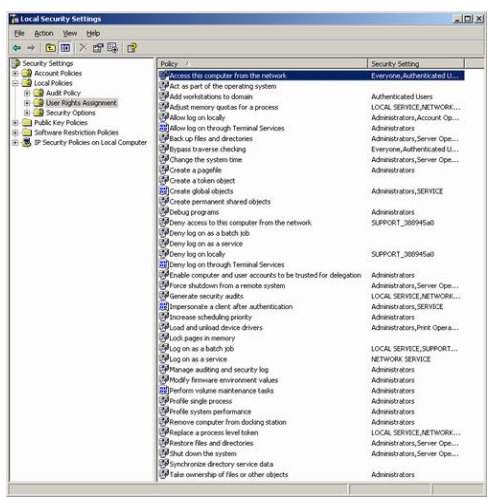

Figure 9. Windows local security policy editor

Another form of privileges exist in Windows, they are called Software Restrictions Policies which enable administrators to control, manage, and disable features of the installed applications on their systems. Figure 10 shows the software restrictions policy editor (gpedit.msc).

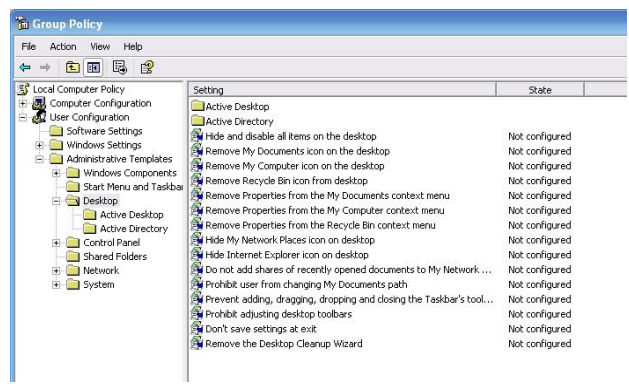

Figure 10. Windows software restrictions policy editor

### Linux

Linux uses Mandatory Access Control (MAC) to enforce privileges. MAC involves aspects that the user cannot control or is not usually allowed to control. Objects are tagged with labels representing the sensitivity of the information contained within. MAC restricts access to objects based on their sensitivity. Subjects need formal clearance (authorization) to access objects [12].

### Head-to-Head Comparison

Linux operating system does not implement the concept of privileges in a separate process as the Windows using LSA; rather, it uses mandatory access control or MAC to restrict access to system objects. Additionally, Linux does not provide the concept of software restrictions; rather, it uses a separate daemon to perform sort of security configuration for specified applications.

## AUDIT

Audit is the process of generating log files as a result of an access check. It shows who accessed what, when, and how [18].

### Windows

The Windows object manager can generate audit events as a result of an access check. Lsass maintains audit information on the local system, and it is configured with the local security policy editor (secpol.msc) [19]. In effect, Lsass sends messages to the SRM (Storage Resource Management) to inform it of the auditing policy at system initialization time and when the policy changes. Lsass is responsible for receiving the generated audit records, editing the audit records, and sending them to the event logger. The event logger then writes the audit record to the security event log. Basically, audit records are put on a queue to be sent to the LSA as they are received, they are not submitted in batches. The audit records are moved from the SRM to the security subsystem in one of two ways. If the audit record is small (less than the maximum LPC message size), it is sent as an LPC (Local Procedure Call) message. The audit records are copied from the address space of the SRM to the address space of the Lsass process. If the audit record is large, the SRM uses shared memory to make the message available to Lsass and simply passes a pointer in an LPC message [14]. Figure 11 illustrates the Windows audit complete mechanism.

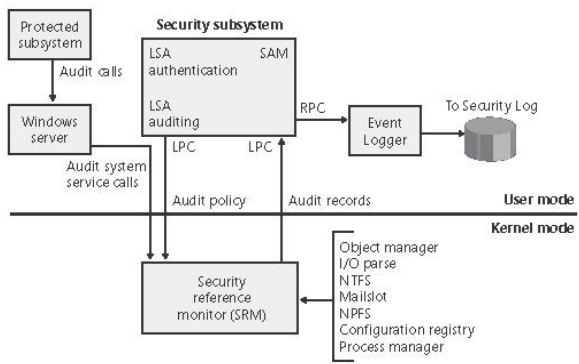

Figure 11. Windows audit mechanism

*Linux*

Traditionally, the Linux kernel and its system processes, store pertinent information in log files, either locally, or centrally on a network server, via the flexible and configurable syslog facility. In addition, many modern Linux systems support a more comprehensive type of auditing known as C2 audit. This is so-name because it fulfills the audit requirements for the TCSEC C2 security level.

*Head-to-Head Comparison*

Both operating systems provide some auditing and logging features through different mechanisms and services. Conversely, Windows provides the System Access Control List or SACL which states what operation over what object should be logged.

## LOGON OR AUTHENTICATION

Logon or authentication is the process by which the identity of a user accessing a system or other source of information is verified. In modern operating systems authentication is commonly done through the use of a username and a password.

*Windows*

Interactive logon occurs through the interaction of the Winlogon logon process (Winlogon.exe), Lsass, one or more authentication packages, and the SAM or Active Directory. Authentication packages are DLLs that perform authentication checks. Kerberos is the Windows authentication package for interactive logon to a domain, and MSV1_0 is the Windows authentication package for interactive logon to a local computer. In fact, Winlogon relies on a Graphical Identification and Authentication (GINA) DLL to obtain a user's account name and password. The default GINA is Msgina (\Windows\System32\Msgina.dll). Msgina presents the standard Windows logon dialog box [14]. Figure 12 shows the various components involved in the Windows logon process.

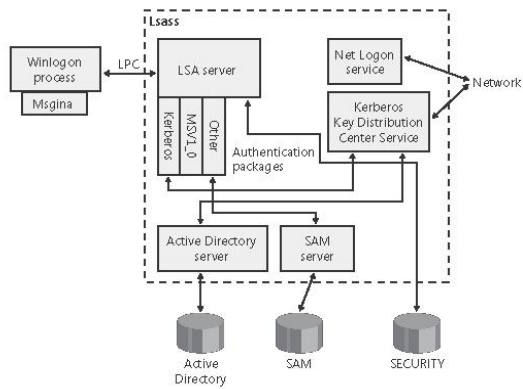

Figure 12. Windows logon process

Windows uses NTLM2 (NT LAN Manager 2) protocol for authentication. The default authentication package that implements this protocol is the MSV1_0 (\Windows\System32\Msv1_0.dll). Below is a list of steps involved in the Windows logon process:

1. A user accesses a client computer and provides a domain name, user name, and password.
2. The client computes a cryptographic hash (HMAC-MD5) of the password and discards the plain text password.
3. The client sends the user name to the server in plaintext
4. The server generates a 16-byte random number, called a challenge and sends it to the client.
5. The client encrypts this challenge using DES with the hash of the user's password (Key) and returns the result to the server. This is called the response.
6. The server sends the following three items to the SAM Server: Username, the challenge sent to the client, and the response received from the client
7. The SAM Server uses the user name to retrieve the hash of the user's password from the SAM database. It uses this password hash (HMAC-MD5) to encrypt using DES the challenge.
8. The SAM Server compares the encrypted challenge it computed (in step 7) to the response computed by the client (in step 5). If they are identical, authentication is successful

*Linux*

On almost all Linux distributions, user information is stored in /etc/passwd, a text file which contains the users' username, their encrypted password, a unique numerical user id (called the UID), a numerical group id (called the gid), an optional comment field (usually containing such items as users' real name, phone number, etc.), their home directory, and their preferred shell [15]. A typical entry in /etc/passwd may look as: *dave:T4xcQ1Nnx7PRE:1000:1000:David Doe,1−700−:/home/dave:/bin/bash*

**1. Shadow Passwords**

The /etc/passwd file, which contains information about all users, including their encrypted passwords, is readable by all users, making it possible for any user to get the encrypted password of everyone on the system. Password cracking programs are widely available. To combat password cracking programs, shadow passwords were developed.

Characteristically, when a system has shadow passwords enabled, the password field in /etc/passwd is replaced by an "x" and the user's real encrypted password is stored in /etc/shadow. Because /etc/shadow is only readable by the root user, malicious users cannot crack their fellow users' passwords. Each entry in /etc/shadow contains the user's login, their encrypted password, and a number of fields relating to password expiration. A typical entry may look as: *dave:/3GJllg1o5477:11209:0:556554:7:::*

Group information is stored in /etc/group. The format is similar to that of /etc/passwd, with the entries containing fields for the group name, password, numerical id (gid), and a comma separated list of group members. An entry in /etc/group may look as: *edu:x:13:student,instructor,assistant*

**2. MD5 Encryption**

Traditionally, Linux passwords were encrypted with the standard crypt() function. As computers grew faster, passwords encrypted with this function became easier to crack. As the internet emerged, tools for distributing the task of password cracking across multiple hosts became available. Many newer distributions ship with the option of encrypting passwords with the stronger MD5 hash algorithm [20]. While MD5 passwords will not eliminate the threat of password cracking, they will make cracking your passwords much more difficult.

*Head-to-Head Comparison*

It is obvious that Windows implements a more secure and elegant mechanism than Linux, though it is more complicated. First, Windows uses the HMAC-MD5 as a hash function which is an enhancement to MD5, the one that Linux uses. Second Windows in some places performs encryption using the Advanced Encryption Standard or AES. Linux does not use any symmetric encryption algorithm. Finally hashed values are stored in Windows in the SAM file; while, in Linux it is stored in a file called /etc/passwd.

**FILE SYSTEM SECURITY**

Computer file systems are employed on data storage devices such as hard disk, to maintain the physical location of the computer files. A file system organizes data in an efficient manner and allows users to create, copy, paste, and delete files.

*Windows*

The NTFS file system is the native file system format of Windows. NTFS includes a number of advanced features, such as file and directory security, disk quotas, file compression, directory-based symbolic links, recoverability, and encryption [21].

Encrypting File System (EFS), which users can use to encrypt sensitive data is a remarkable feature of the NTFS file system. The operation of EFS is completely transparent to applications, which means that file data are automatically decrypted when an application running in the account of a user that is authorized to view the data and automatically encrypted when an authorized application changes the data. Below is a list of steps involved in the encryption of a file under EFS:

1. When a file is encrypted, the file system generates a random number file encryption key (FEK). FEK is used to encrypt the file's contents using Advanced Encryption Standard (AES).
2. FEK is stored with the file but encrypted with the user's EFS public key by using the RSA.
3. After EFS completes these steps, the file is secure. Other users can't decrypt the data without the file's decrypted FEK, and they can't decrypt the FEK without the private key.

*Linux*

File system in Windows has more security features than the one in Linux; however, Linux distributions are starting to use Extended Access Control Lists (EACL) as a part of their file system, bringing it more on par with NTFS. These EACL in Linux systems are defined by the file mode. The file mode comprises nine flag bits that determine access permissions of a file. This mechanism allows defining access permissions for three classes of users: the file owner, the file group, and others. These permissions can be used to prevent a certain user belonging to a certain class from accessing the file data of another user that he is not authorized to access [22].

*Head-to-Head Comparison*

Unlike Windows, Linux does not provide file encryption as part of its native file system; however, Linux through third party tools can achieve some level of encryption over files and directories such as the GnuPG using the gpg command.

**CONCLUSIONS**

This paper discussed the different security aspects of the most two successful commercial operating systems, Microsoft Windows and Linux. The various security features, designs, and components of the two OS were covered extensively, showing the key similarities and differences between them. In fact, both OS have a lot of common security concepts and mechanisms, though sometimes implemented differently, such as object identification, user authentication, access token, access control lists, and others. What differ between them are few attributes such as file system encryption and software privileges which Windows have and Linux don't, and shadow password which Linux has and Windows doesn't. It was obvious from the conducted analysis that Windows OS incorporates more of its security components within its kernel; while, Linux counts more on user-mode processes. Moreover, Windows uses complicated features such as audit; while, Linux uses less intricate, yet efficient, log files with encryption. Overall, both operating systems provide comparatively adequate multi-level security technologies making them both certified as trusted operating systems that can cope with hostile situations and attacks, and provide a secure environment for computer users and their applications.